\documentstyle[prb,aps]{revtex}        
\input epsf
\tighten
\begin{document}

\twocolumn[\hsize\textwidth\columnwidth\hsize\csname @twocolumnfalse\endcsname
\title{Reentrant behaviour and universality in the Anderson transition}
\author{ S. L. A. de Queiroz\footnote{Electronic address:
sldq@if.ufrj.br}} 
\address{
Instituto de F\'\i sica, Universidade Federal do Rio de Janeiro,\\
Caixa Postal 68528, 21945-970  Rio de Janeiro RJ, Brazil}
\date{\today}
\maketitle
\begin{abstract}
The three-dimensional Anderson model with a
rectangular distribution of site disorder displays two distinct
localization-delocalization transitions, against varying disorder
intensity, for a relatively narrow range of Fermi energies. Such
transitions are studied through the calculation of
localization lengths 
of quasi-- one-dimensional systems  by transfer-matrix
methods, and their analysis by finite-size scaling techniques.
For the transition at higher disorder we find the localization-length
exponent $\nu=1.60(5)$ and the limiting scaled
localization-length amplitude $\Lambda_0=0.57(1)$, strongly suggesting
universality with the transition at the band centre, for which
currently accepted values are $\nu=1.57(2)$ and $\Lambda_0=0.576(2)$. 
For the lower (reentrant) transition, we estimate $\nu=1.55(15)$ and
$\Lambda_0=0.55(5)$, still compatible with universality but much less
precise, partly owing to significant finite-size
corrections.

\end{abstract}

\pacs{PACS numbers:  71.30.+h, 71.23.-k, 72.15.-v,72.15.Rn}
\twocolumn
\narrowtext
\vskip0.5pc]

\section{Introduction}
\label{intro}
According to Anderson's original scenario~\cite{and58}, if one starts from
a translationally invariant 
model Hamiltonian for noninteracting electrons and considers increasing
quenched disorder (e.g. by assigning random self-energies to lattice 
sites), a critical threshold is reached at which quantum
interference inhibits propagation, and the system undergoes a localization
transition, from a diffusive (metallic) phase to an insulating
(localized) regime.  
For the past twenty years, the framework for studies of critical
properties of the Anderson transition has mostly been provided by the
scaling theory of localization~\cite{gof4,mkk93}. Numerical studies have
been progressively 
refined~\cite{ps81,mkk81,mkk83,bkmk85,bsk87,cbs90,mk94,so99,osk99,sok00}, 
so that accurate estimates exist for several quantities of interest.\par
In the site-disordered version which will concern us here, the problem is
usually formulated in terms of a tight-binding Hamiltonian on a regular
lattice,
\begin{equation}
{\cal H} = \sum_i \varepsilon_i|i\rangle\,\langle i| + 
V \sum_{\langle i,j\rangle}|i\rangle\,\langle j| \ \ \ ,
\label{eq:ham}
\end{equation}
where the site self-energies $\varepsilon_i$ are independent,
identically distributed random variables obeying a specified
distribution,
$\langle i,j\rangle$ denotes nearest neighbours and the
energy scale is set by the hopping matrix element, $V \equiv 1$.
The intensity of disorder is generically represented by a
width, $W$, of the self-energy probability distribution. \par
On a simple cubic lattice, for the non-random case (all $\varepsilon_i = 
\varepsilon _0 \equiv 0$, for convenience) the Schr\"odinger equation
${\cal H}\,|\psi\rangle =
E\,|\psi\rangle$ yields Bloch (propagating) states for $-6 \leq E \leq 6$
and no solutions outside this energy range. In many studies of the
localization
transition~\cite{ps81,mkk81,mkk83,mk94,so99,osk99,sok00} 
one looks for destruction of transmission at the band centre, that is,
only the behaviour of states at $E=0$ is considered. 
Since that is as far as possible from the unperturbed band edges,
one is maximising the direct influence of quenched disorder which,
coupled with the ever-present quantum interference effects, induces
Anderson localization  
(compare this with the purely quantum mechanism responsible for the
very existence of a band of allowed states, and forbidden ones outside it, 
even in a homogeneous system).\par
Further studies have investigated the $E \neq 0$ region, attempting to
map the mobility edge trajectory in the
$E-W$ plane~\cite{bkmk85,bsk87,cbs90,gs95}. Along this latter
line, reentrant behaviour was found for
certain distributions of site self-energies, in a narrow range of
energies above the unperturbed band edge~\cite{bkmk85,bsk87,gs95}: 
at fixed $E$, as $W$ decreases from some large initial value the system
first displays an upper transition from insulator to metal at $W \to
W_u^+$, much the same as at $E=0$~; then, upon further reduction of
disorder a second, lower transition takes
place, back towards an insulating phase as $W \to W_l^+$. The mechanisms
underlying each of these transitions are qualitatively understood, and 
fundamentally distinct~\cite{bkmk85,bsk87}: while the onset of extended
states as
$W \to W_u^+$ is due to diminishing quantum interference, at $W \to W_l^-$
it is the increasing overlap of (and consequent tunnelling
between) localized states that eventually
produces delocalized behaviour. Reentrant regions were found for
rectangular ($P(\varepsilon_i) = {\rm constant,}\ |\varepsilon_i| \leq
W/2)$ and, to a lesser extent, for Gaussian ($P(\varepsilon_i) \propto
\exp(-\varepsilon_i^2/2W^2)$) distributions, and not at all for a
Lorentzian one ($P(\varepsilon_i) \propto
W/\left[\varepsilon_i^2 +W^2\right]$). This is consistent with
the idea that, in order for localized states to coalesce into one
extended wavefunction (the mechanism underlying the lower transition),
their respective energies must not differ by
much, thus enabling their mutual resonance. The narrower the distribution
of self-energies (preferably with a sharp cutoff, as in the rectangular
case), such situation is more likely to be found. For a Lorentzian
shape, where already the second moment diverges, chances of
overlap are negligible. \par
Though significant advances have been made recently regarding
properties at the band centre~\cite{mk94,so99,osk99,sok00},
such as universality
(with respect to varying disorder distributions), corrections to
scaling  and accurate numerical estimates of critical quantities, 
a similar quantitative 
understanding of behaviour elsewhere on the phase diagram appears to be
lacking. Here we set the model parameters in order to enable a study
of the reentrant region, and probe both upper and lower transitions;
our main goal is to find what universality class(es) they belong to.
Our numerical procedures and some concurring technical aspects are
discussed in Section ~\ref{secII}; the upper (high disorder)
and  lower (reentrant) transitions are analyzed respectively in Sections
~\ref{secIII}  and ~\ref{secIV}; conclusions and final remarks are
given in Section~\ref{conc}.

\section{Model details and Calculational method}
\label{secII}

We consider the Hamiltonian given in Eq.\ (\ref{eq:ham}), on a
simple cubic lattice, with a rectangular distribution of site
self-energies:
\begin{equation}
P (\varepsilon_i) = \cases{{\rm constant} \quad -W/2 \leq \varepsilon_i
\leq +W/2 \cr
                           0\qquad\qquad\quad {\rm otherwise}}
\label{eq:prob}
\end{equation} 
\begin{figure}
\epsfxsize=8.4cm
\begin{center}
\leavevmode
\epsffile{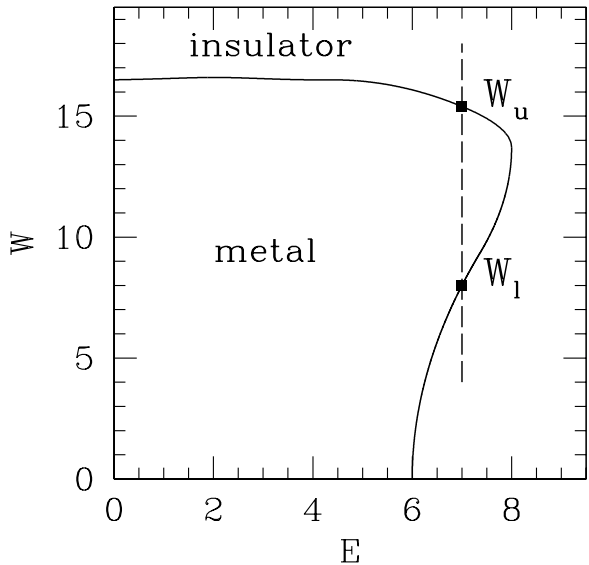}
\caption{Schematic phase diagram for rectangular disorder distribution
(after Ref. 8, see their figure 1). Dashed
line at $E=7$ shows region investigated here, for which the lower and 
upper transitions are located  at $W_l \simeq 8$,
$W_u \simeq 15$, respectively.
}
\label{fig:qpd}
\end{center}
\end{figure}

For such distribution the mobility edge trajectory, as found in
Ref.~\onlinecite{bsk87}, is semi-quantitatively depicted in
Figure~\ref{fig:qpd}. From inspection of the Figure, we decided for fixing
the energy at $E=7$ and scanning $W$. This way two well-separated
transitions (respectively at  $W_u \simeq 8$,
$W_l \simeq 15$) are expected to be present, so presumably one can
focus on their individual properties one at a time.\par
We have considered quasi-- one-dimensional systems, ``bars'' of
cross-section $L \times L$ sites, with periodic boundary
conditions across, and very long length $N$. We have used
$L=4$, 6, 8, 10, 12 and 14; as explained below, the values of $N$
were determined by numerical convergence criteria, $10^5$ being
typical.\par
Following well-known
procedures introduced by Pichard and Sarma~\cite{ps81} (see also 
Refs. \onlinecite{mkk83,mk94} for detailed descriptions), we
considered the transfer matrix connecting matrix elements of the
Hamiltonian, Eq.\ (\ref{eq:ham}), on consecutive cross-sections,
and iterated it along the
``infinite'' direction, thus extracting the corresponding Lyapunov
characteristic exponents.
Of these, the most relevant here is the one with smallest modulus,
which gives the largest localization length $\lambda_L(E,W)$ 
(slowest decay along the bar of transverse dimension $L$, at given
$W$, for an electronic wave function with energy $E$).\par
According to finite-size scaling (FSS)~\cite{bar83},
the non-diverging quasi-- one-dimensional lengths $\lambda_L$
scale linearly with $L$ (apart from corrections to scaling, to be
discussed below) at the critical point of the three-dimensional
system. Therefore, plots of the scaled localization lengths $\Lambda_L
\equiv \lambda_L/L$ (e.g. against
$W$, for fixed $E$) for different values of $L$ must cross, and they
should do so twice if there are two distinct second-order transitions.
Leaving aside, for the moment, the question of whether this is consistent
with the single-parameter theory of localization (see Ref. 
\onlinecite{bsk87}), one is provided with a rough-and-ready criterion to
locate the expected critical points.\par
In order to produce reliable numerical estimates from the conceptual
framework described thus far, two points must be carefully considered.
First, though the convergence of finite-$N$ estimates of Lyapunov
exponents towards their asymptotic values is guaranteed in
principle~\cite{ose68,ranmat}, it may be hard to decide when it has
taken place to a desired accuracy. As noted in Ref. \onlinecite{mkk83},
``for such very long strips the variation of $\lambda$ [as a function
of $N$] from step to step is very small due to
averaging, and this can lead to a false impression of convergence''.
Accordingly, we have kept track of the variance among
estimates~\cite{mkk83}, in
the following way. Each $10^3$ steps an estimate of $\lambda$ 
(taking into account all iterations from the origin up to the sampling
point) would be produced; first, twenty consecutive estimates would
be accumulated, and the average and RMS deviation among them
calculated. At the
end of each subsequent $10^3$ steps, a new average and deviation
would be calculated, of the {\em most recent} twenty estimates. Therefore,
in this moving average procedure, one keeps track of fluctuations
along the latest $20,000$ iterations. It was found that, in order to
reach $0.1\%$ fractional deviation one typically needs $80,000-130,000$
iterations (except for low disorder, $W \lesssim 6.5$, where similar
figures are achieved within $\sim 50,000$ steps or fewer). Strictly
speaking, this is no guaranteee that the corresponding final estimate will
be {\em accurate} within $0.1\%$, as a very slow drift may still be masked
underneath apparently random fluctuations. In Ref. \onlinecite{so99},
$10^6-10^7$  iterations at the band centre are quoted as necessary in
order to give
$0.1\%-0.05\%$ accuracy, although the latter term is not assigned a
specific
operational definition; in the closely related Green's function
calculations of Ref. \onlinecite{bsk87}, $1\%$ accuracy is claimed for
$\sim 40,000-70,000$ steps. Having this in mind, it seems that our
current definition of $0.1\%$ relative fluctuation warrants perhaps
$0.3\%-0.4\%$ accuracy for the localization length, at best.
We used $0.1\%$ relative fluctuations for all $L=4-10$ data, and for
selected $L=12-14$ points; in the latter case, intermediate--$W$ data were
collected with $0.5\%-1\%$ fluctuations and checked for smooth
variation
against the more accurate ones. 
As shown below, this will not be     
the most important source of errors in the present case.\par
Secondly, in general there may be systematic distortions to the linear,
single-scaling parameter picture, caused by nonlinearity of scaling
fields and/or irrelevant variables~\cite{cardy}. A detailed study
of such effects for the Anderson transition at the band centre has
been undertaken recently~\cite{so99}. The result was that such
corrections are of sizable importance in the case: estimates of
critical quantities produced when they are taken into account tend
to be more consistently reliable and accurate than if they are
disregarded.
Following the notation of Ref.~\onlinecite{so99}, near the critical
disorder $W_c$, with $w \equiv (W_c -W)/W_c$, the scaled localization
length $\Lambda_L$ must vary as:
\begin{equation}
\Lambda_L= F(\chi L^{1/\nu},\psi L^y)\ ,
\label{eq:deflam}
\end{equation}
where $\chi$ is the relevant variable, $\psi$ is the leading irrelevant
variable, the exponent $\nu$ characterizes the divergence of the
thermodinamic ($L \to \infty$) localization length via $\lambda_{\infty}
\propto w^{-\nu}$, and $y <0$ is the leading irrelevant
exponent~\cite{bar83,cardy}.
Since there is no transition for finite $L$, one assumes 
\begin{equation}
\Lambda_L = \sum_{n=0}^{n_I} \psi^n L^{ny} F_n(\chi L^{1/\nu})\ ;
\label{eq:deflam2}
\end{equation}
further, since (for finite $L$) $\chi$ and 
$\psi$  also are smooth functions of $w$, one expands both in power
series as well;
linearity is recovered if $\chi(w)=w$ (as $\chi(0)=0$ must hold),
$\psi(w)=\psi$ (constant), and only the lowest powers  are
kept ($n_I=1$) in the Taylor series for $\Lambda_L$:
\begin{equation}
\Lambda_L = F_0(w L^{1/\nu}) + \psi L^y  F_1(w L^{1/\nu})\ .
\label{eq:deflam3}
\end{equation} 
Nonlinearity (both in relevant and irrelevant variables)
and higher-order corrections associated to the irrelevant
variable may be incorporated by keeping higher-order terms in the
respective expansions. This is done at the expense of having a larger
number of parameters to fit, and must be accompanied by the corresponding
quality of numerical data. Here, we found that the quality of our data
was generally
consistent with the linearized form, Eq.~(\ref{eq:deflam3}). 

\section{upper transition}
\label{secIII} 

\begin{figure}
\epsfxsize=8.4cm
\begin{center}
\leavevmode
\epsffile{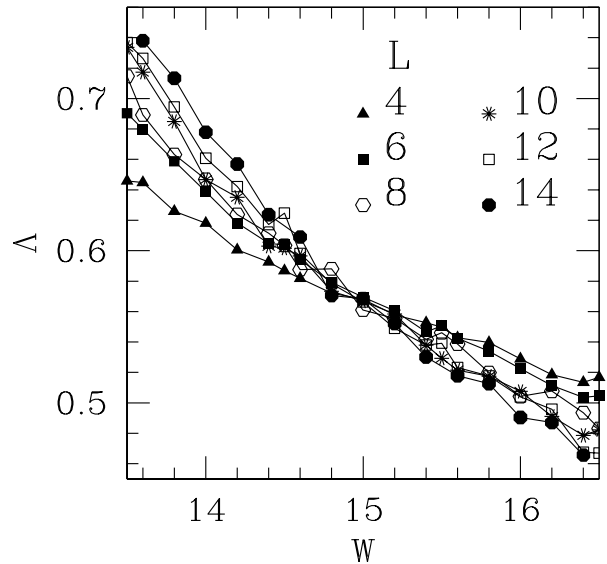}
\caption{Raw data for scaled localization lengths near the upper
transition. Relative fluctuations, as defined in the text, are of order of
symbol sizes or smaller.
}
\label{fig:rawu}
\end{center}
\end{figure}

\begin{figure}
\epsfxsize=8.4cm
\begin{center}
\leavevmode
\epsffile{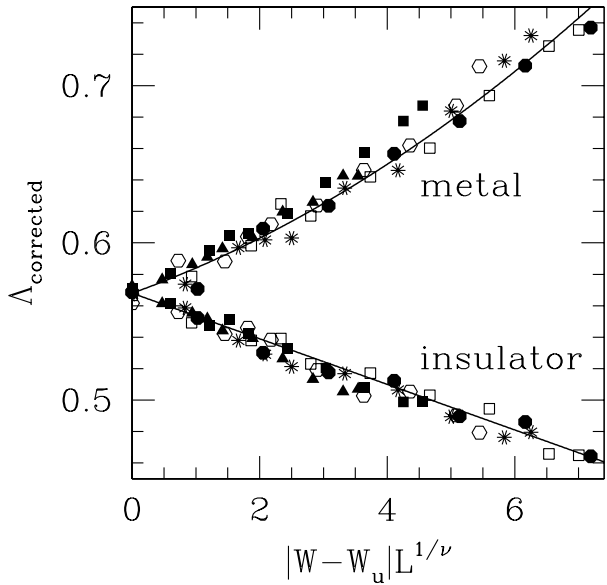}
\caption{ Scaling plot near the upper transition. Here, $W_u=15.0$,
$\nu=1.61$, $\psi=-0.21$, $y=-2.8$, $a=-0.21$, $b=-0.2$ (see 
Eq.~(\protect{\ref{eq:lambcor}})). Lines are guides to the eye, starting
at $\Lambda_0=0.568$ on vertical axis. Key to symbols is the same as in 
Fig.~\protect{\ref{fig:rawu}}.
}
\label{fig:uts}
\end{center}
\end{figure}

In Figure~\ref{fig:rawu} we display the raw data for scaled localization
lengths near the upper transition, whose approximate location is estimated
from Ref.~\onlinecite{bsk87} (see also Figure~\ref{fig:qpd} above). 
Though data for each $L$ behave somewhat irregularly, there is an
unmistakeable crossing
of curves, roughly at the centre of the diagram.   
Following along the lines proposed in Ref.~\onlinecite{so99}, we attempted
to fit assorted corrections to scaling. We found that, to be consistent
with the range of variation and accuracy level of our data, the most
sensible choice
was (1) not to take nonlinearities of scaling variables into account, and
(2) keep only up to quadratic terms in the Taylor expansion of
$F_1(w L^{1/\nu})$ in
Eq.~(\ref{eq:deflam3}) above. Thus we plotted
\begin{equation}
\Lambda_{\rm corrected} \equiv \Lambda_L - \psi L^y\left(1 + ax
+bx^2\right)
\label{eq:lambcor}
\end{equation}
against $x \equiv \left(W-W_u\right) L^{1/\nu}$ . This gave us six
parameters to fit:
$W_u$, $\nu$, $\psi$, $y$, $a$ and $b$.     
We started from reasonable guesses for some of these quantities
(e.g. $W_u \simeq 15$, $|\psi| < 1$); also, since this transition is
 driven by the same mechanism  as that at the band centre,
they might plausibly be in
the same universality class, so $\nu \simeq
1.4-1.7$  seemed worth investigating (recall $\nu=1.57(2)$ from 
Ref.~\onlinecite{so99}). With a few similar aditional assumptions, 
we checked the variation of $\chi^2$ against changes
in the fitting parameters~\cite{numrec}; the plot displayed in
Figure~\ref{fig:uts}
is representative of the best sort thus obtained. Though its
(unweighted) $\chi^2=0.018$ for 110 data
does not mean much by itself, it will be useful for comparison with
the corresponding one at the lower transition, in Section~\ref{secIV}.
Corrections to scaling are of order up to $5\%$ for $L=4$ and decrease
steeply, owing to the large value of $|y|$ (similar to the picture at the
band centre, see Table II of Ref.~\onlinecite{so99}), to
around $0.2\%$ for $L=14$. The latter figure is close to the intrinsic
numerical accuracy of raw data, so it is fair to say that corrections only
play a significant role for the smaller sizes, say up to $L=8-10$.
Our plots turned out
to be quite sensitive against changes in $\nu$ of order $3-4\%$ around
the best-fitting value $\simeq 1.6$;
therefore,
the initial hypothesis of universality, as incorporated in the fitting
procedure, appears to be rather consistent.\par
On the other hand,  the limiting amplitude,
$\Lambda_0 \equiv \lim_{L\to \infty,W=W_c} \Lambda_L$, 
given on scaling plots such as Figure~\ref{fig:uts} by the
intersection with the vertical axis, was found to be
remarkably robust: we quote $\Lambda_0 = 0.57(1)$  for all
reasonably behaved choices of parameters. This is in very good 
agreement with $\Lambda_0 = 0.576(2)$ , a value found to be
universal with respect to the form of disorder
distribution at the band centre~\cite{so99}. 
\begin{table}
\caption{Estimates of critical quantities at upper transition.
Data for the band centre from Ref. 11.
Comparison of values for $W_c$ is not pertinent. 
}
\vskip 0.1cm
 \halign to \hsize{\hfil#\hfil&\hfil#\hfil&\hfil#\hfil&
\hfil#\hfil&\hfil#\hfil\cr
     \ \      &\ $W_c$ &\ $\Lambda_0$ &\ $\nu$ &\ $y$ \cr
 Present work &\ $15.0(1)$ &\ $0.57(1)$ &\  $1.60(5)$ &\ $-2.8(5)$ \cr
 Band centre &\ $16.54(1)$ &\ $0.576(2)$ &\ $1.57(2)$ &\ $-2.8(5)$ \cr}
\label{table1}
\end{table}
\noindent
Note also that $\Lambda_0$ does not directly
depend on the initial choice of fitting parameters; instead, it is
obtained at the end of the procedure, so the likelihood of this being
a  biased result is smaller than e.g. in the case of $\nu$.
Universal critical amplitudes play an important role in the statistical
mechanics of critical phenomena~\cite{cardy,pha91}, so the
present numerical
agreement  must be regarded as strong evidence in favour
of universality of the transition (at least along the upper,
non-reentrant, portion of the mobility edge, including the band
centre).\par
To our knowledge, the only existing calculation of exponents along the  
mobility edge is that of Ref.~\onlinecite{cbs90} for Gaussian disorder
(see their Table IV); however, their estimates are far too scattered, and
no clear conclusion can be drawn from them.\par
In Table ~\ref{table1} we summarize our estimates for critical quantities
at the upper transition; for comparison, corresponding values at the
band centre are quoted. Error bars
for the present work's results are somewhat arbitrary, but certainly
encompass all well-behaved scaling plots. The exact coincidence of the
estimate for
$y$ with that of Ref.~\onlinecite{so99} must be regarded as
fortuitous. The overall conclusion is that
both transitions are in the same universality class.

\section{lower (reentrant) transition}
\label{secIV} 
\begin{figure}
\epsfxsize=8.4cm
\begin{center}
\leavevmode
\epsffile{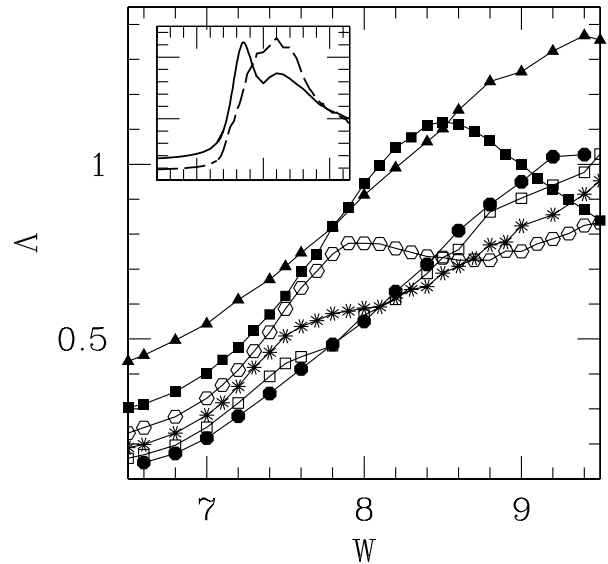}
\caption{Raw data for scaled localization lengths near the lower
transition. Key to symbols is the same as in
Fig.~\protect{\ref{fig:rawu}}. Inset: raw data for $L=6$ (full line)
and $L=12$ (dashed line), $2 \leq W \leq 16.5$.
}
\label{fig:rawl}
\end{center}
\end{figure}

In Figure~\ref{fig:rawl} are displayed the raw data for scaled
localization lengths near the lower transition, whose approximate location
is again estimated
from Ref.~\onlinecite{bsk87}.
Data here behave rather differently from the
upper transition. Non-monotonicity has been found earlier in this
region; see data of Ref.~\onlinecite{bsk87} for $E=6.6$, $6.8$, $W=8$,
$L=6-10$. A proper perspective is gained by looking at a broader
range of $W$--variation, including both transitions: see the inset
of Figure~\ref{fig:rawl}, where only $L=6$ and $12$ data are displayed, 
for clarity. The maxima shown in the main 
diagram are in fact shoulders, attributable to irrelevant scaling
variables, much in the same way as the corrections discussed in
Ref.~\onlinecite{so99} and Section~\ref{secIII} above. They must
vanish for increasing $L$, and are distinct from the bona fide maxima,
expected
to be exhibited by finite-size data deep inside the conducting phase,
somewhere between the two transitions. Indeed, the inset clearly shows
that for $L=6$ the former effect is stronger than the latter, while for
$L=12$ the trend has very much been  reversed. For $L=4$ (not shown)
shoulder and maximum (if any) merge, owing to
finite-size broadening of the former.\par
\begin{figure}
\epsfxsize=8.4cm
\begin{center}
\leavevmode
\epsffile{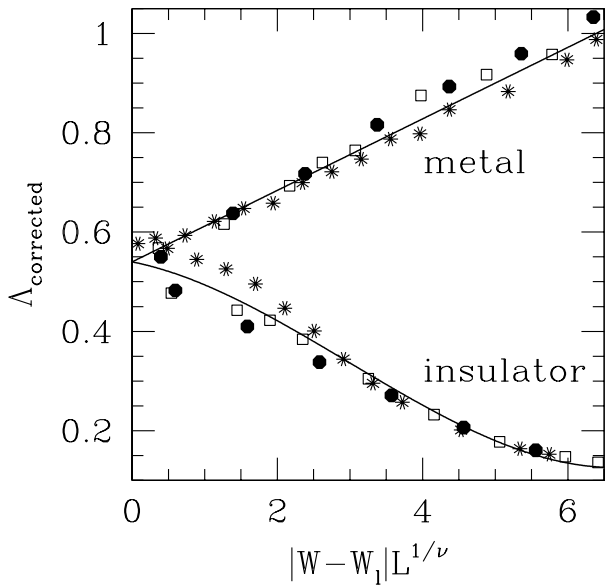}
\caption{ Scaling plot near the lower transition. Here, $W_l=7.9$,
$\nu=1.65$, $\psi=4.5$, $y=-3.2$, $a=-2.07$, $b=0$ (see 
Eq.~(\protect{\ref{eq:lambcor}})). Lines are guides to the eye, starting
at $\Lambda_0=0.54$ on vertical axis. Key to symbols is the same as in 
Fig.~\protect{\ref{fig:rawu}}; only data for $L=10-14$ are used.
}
\label{fig:lts}
\end{center}
\end{figure}

The magnitude of distortions thus displayed makes it hard to 
account for them within a Taylor expansion framework, with a
realistic number of parameters. Attempts to subtract the shoulders
by postulating an {\it ad hoc} gaussian shape were not successful.
We then decided to discard $L=4-8$ data, and try a fitting procedure
similar to that given in Eq.~(\ref{eq:lambcor}) above, using only data
for $L=10$, $12$, and $14$. Figure~\ref{fig:lts} shows the results of the
best fit, with a set of parameters for which the
(unweighted) $\chi^2=0.10$  for 61 data.\par

One sees that the shoulder of $L=10$ data has been reduced, but not
totally eliminated; also, the overall quality of data collapse is inferior
to, and the amplitudes of corrections larger than, the corresponding
ones  at the upper transition. However, the
values  $\nu = 1.65$ and $\Lambda_0=0.54$ used in  Fig.~\ref{fig:lts}
are within $6\%$ of those at the band centre. Bearing in mind the
numerous  assumptions made along the way, assigning error bars
to these numbers would be somewhat risky. We have found that plots 
using $\nu$ just about anywhere in the $1.4-1.7$ range do not seem
obviously much worse than the one displayed above; as regards $\Lambda_0$,
anything between $0.5$ and $0.6$ seems possible, provided that one keeps
$W_l$ in the range $7.8-8.2$. Therefore, the proper conclusion is that
our data
are not inconsistent  with the lower transition being in the same
universality
class as that at the band centre (and, presumably, along the upper
portion of the mobility edge).

\section{conclusions}
\label{conc}
We have found strong evidence that the transition along the upper portion 
of the mobility edge is in the same universality class as that at the band
centre. We quote $\nu=1.60(5)$ and $\Lambda_0=0.57(1)$, to be compared
respectively with $\nu=1.57(2)$ and $\Lambda_0=0.576(2)$ of
Ref.~\onlinecite{so99}.\par
Our results for the lower (reentrant) transition are
more mixed, though they certainly are compatible with universality: we
give $\nu=1.55(15)$ and $\Lambda_0=0.55(5)$, from a qualitative visual
analysis of trial scaling plots. One likely source for this would be
the significant presence of finite-size corrections, linked to
irrelevant scaling variables and/or nonlinearities in the scaling
fields; so far, we have not been able to account for the bulk of such
effects within a Taylor-expansion picture similar to that successfully
employed, both at the band centre in Ref.~\onlinecite{so99} and at the
upper transition in the present work.\par
Another, more fundamental, reason for such irregularities would be the
failure of one-parameter scaling theory to describe the lower transition
at all. It is argued in Ref.~\onlinecite{bsk87} that ``the critical
behaviour should be {\em independent} of whether the energy $E$, the
disorder $W$, or some combination of both, is taken as the scaling
variable ... this is not generally true [e.g., in the
present case] because the system shows {\em two} metal-nonmetal 
transitions when the disorder is changed for a fixed energy''.
It is also true that the physical mechanisms driving the lower
transition (quantum tunnelling between tail states) and the upper
one (quantum interference between essentially delocalized wave
functions) are, in principle, distinct. Thus perhaps the one relevant
parameter that represents the latter physical effect cannot do so for
the former.\par
On the other hand, reasonably good scaling plots have been
found here, by implicitly assuming the validity of one-parameter scaling,
with corresponding numerical values of critical quantities which are
not inconsistent with universality. This hints that the picture along 
both the upper and reentrant parts of the
mobility edge may, in fact, be rather simple: critical behaviour would be
dominated by a single fixed point, located at $E=0$.\par
More work, numerical and theoretical, is clearly necessary in order to 
settle the issue. It is also expected that experimentalists could be
attracted into the discussion;  metallic
samples with suitably-located Fermi energies might conceivably
be made to scan reentrant regions, by varying dopant concentrations.

\acknowledgments
The author thanks Belita Koiller and J. A. Castro for interesting
discussions,
and Brazilian agencies CNPq (grant \# 30.1692/81.5), FAPERJ (grants
\# E26--171.447/97 and \# E26--151.869/2000) and FUJB-UFRJ 
for financial support.

\end{document}